**Epidemiology of exposure to mixtures: we can't be casual about causality when using or testing methods**

Thomas F. Webster[1] and Marc G. Weisskopf[2]

[1] Department of Environmental Health, Boston University School of Public Health, Boston, MA, USA

[2] Department of Environmental Health, Harvard T.H. Chan School of Public Health, Boston, MA, USA

Corresponding author:

Dr. Thomas F. Webster

ORCiD: 0000-0003-4896-9323

Department of Environmental Health (T4W)

Boston University School of Public Health

715 Albany St

Boston, MA 02118 USA

email: twebster@bu.edu

**Abstract**

**Background:** There is increasing interest in approaches for analyzing the effect of exposure mixtures on health. A key issue is how to simultaneously analyze often highly collinear components of the mixture, which can create problems such as confounding by co-exposure and co-exposure amplification bias (CAB). Evaluation of novel mixtures methods, typically using synthetic data, is critical to their ultimate utility.

**Objectives:** This paper aims to answer two questions. How do causal models inform the interpretation of statistical models and the creation of synthetic data used to test them? Are novel mixtures methods susceptible to CAB?

**Methods:** We use directed acyclic graphs (DAGs) and linear models to derive closed form solutions for model parameters to examine how underlying causal assumptions affect the interpretation of model results.

**Results:** The same beta coefficients estimated by a statistical model can have different interpretations depending on the assumed causal structure. Similarly, the method used to simulate data can have implications for the underlying DAG (and vice versa), and therefore the identification of the parameter being estimated with an analytic approach. We demonstrate that methods that can reproduce results of linear regression, such as Bayesian kernel machine regression and the new quantile g-computation approach, will be subject to CAB. However, under some conditions, estimates of an overall effect of the mixture is not subject to CAB and even has reduced uncontrolled bias.

**Discussion:** Just as DAGs encode *a priori* subject matter knowledge allowing identification of variable control needed to block analytic bias, we recommend explicitly identifying DAGs underlying synthetic data created to test statistical mixtures approaches. Estimates of the total effect of a mixture is an important but relatively underexplored topic that warrants further investigation.

**Introduction**

It has long been recognized that humans and ecosystems are exposed to multiple chemical and non-chemical agents that may affect health. While environmental epidemiologists tend to analyze exposures one at a time, methods for examining exposure to mixtures of exposures have recently received increased attention, including conferences and grants comparing approaches (Carlin et al 2013, Taylor et al. 2016, PRIME). Questions of interest to mixtures epidemiology include selection of exposure variables that contribute to the outcome, interactions, joint effect of the mixture as a whole and construction of exposure summary measures. Important issues that come up in this context include collinearity and confounding by co-exposures (Braun et al 2016).

Causality is of great concern to epidemiology, but there has been a revolution in causal methods in the last decade or two. The use of directed acyclic graphs (DAGs) is one such powerful method (e.g., Hernan and Robins 2020). Causal methods are now widely seen as critical to the design and analysis of etiologic epidemiologic studies (exempt from this rule are descriptive studies or purely predictive models, as long as it is clear that their purpose is not etiologic investigation). In particular, one needs to have a causal model in mind when interpreting epidemiologic results.

However, causality has so far received relatively little attention in mixtures epidemiology (Weisskopf et al. 2018, Keil et al 2020). Analogously to the idea that causal models are necessary to evaluate possible confounding as well as collider bias (Hernán et al 2002), we will also argue here that causal models are necessary when testing and comparing methods for mixtures epidemiology. This includes construction and analysis of synthetic data that are often used for evaluation of approaches without closed form solutions, as is true for many novel methods.

Correlation between exposures is common and creates both epidemiologic and statistical issues. In this paper we explore the intersection of mixtures epidemiology and causality by examining two questions that relate to collinearity. 1) How is a given mixtures epidemiology method affected by increasing collinearity? In particular, how should we create synthetic data to answer this question? 2) Are newer mixtures methods, such as Bayesian kernel machine regression (bkmr) (Bobb et al 2015) and quantile g-computation (qgcomp) (Keil et al 2020), susceptible to co-exposure amplification bias (CAB), which we have previously described as a potential problem when examining multiple correlated exposures (Weisskopf et al. 2018)? Before addressing these questions, we will briefly review causal models and mixtures epidemiology.

## 1. Background: Causal models and mixtures epidemiology

To keep things simple, we will restrict ourselves to the situation of one outcome, two exposures and a limited number of confounders. Exposure to mixtures can obviously be far more complex (e.g., Weisskopf et al. 2018). While this represents a very simple mixtures problem, interpretation can still be complicated. For now, let's consider two possible causal models:

### 1a. Confounding by co-exposure

As shown in the DAG in Figure 1a, assume that the two exposures, X1 and X2, are correlated due to an unknown or unmeasured cause U, e.g., a common source. The DAG, if true, represents the causal connections between variables (the arrows). For a review of DAGs and related terminology see (Glymour and Greenland 2008; Hernan and Robins 2020; Greenland et al. 1999). Here, both X1 and X2 are causally associated with the outcome Y (Throughout this paper we assume no effect measure modification of these relationships or interactions, which are typically not reflected in DAGs). Suppose that we are interested in the contribution of each mixture component to the outcome. The crude X1-Y association is confounded by X2 because in addition to the causal pathway from X1 to Y, there is also an open pathway from X1 to Y via X2. Similarly, the X2-Y association is confounded by X1. We call this problem confounding by co-exposure. A typical solution is to mutually adjust for both X1 and X2, e.g., put both in a regression model. For this DAG, mutual adjustment produces unbiased results. The X1-Y association is unbiased because adjusting for X2 blocks the pathway from X1 to Y via X2, and vice versa. These conclusions do not depend on the mathematical form of the relationships underlying the links between variables (Greenland et al 1999).

The magnitude of the crude and mutually adjusted effect estimates have straightforward closed-form solutions when the variables are continuous, the associations are linear, and we apply linear regression. Let $b_1$ be the causal coefficient linking X1 to Y, i.e., the increase in Y caused by a one unit increase in X1, and similarly for $b_2$ linking X2 to Y. For simplicity, we'll assume that the variables are standardized (centered with unit variances). Let $\rho$ be the correlation coefficient for X1 and X2 (combining the two causal coefficients $b_3$ and $b_4$ linking U to X1 and X2. Note that the other pathway between X1 and X2 via Y is blocked because Y is a collider). Table 1 shows the crude and adjusted regression coefficients (see Weisskopf et al. 2018 for derivations).

1b. Co-exposure amplification bias

Now let's examine another possible DAG for X1, X2 and Y. Figure 1b describes a situation called co-exposure amplification bias (CAB) (Weisskopf et al. 2018), a variation of z-amplification bias (e.g., Pearl 2010). X1 and X2 are correlated as before because of U, but only X1 is causally associated with Y. Suppose that there is also an unknown (or unmeasured) variable U' that confounds the crude X1-Y association. For example, this type of situation might occur when X1 is a biomarker measured in serum and U' is a physiological factor that affects both the biomarker X1 and the outcome. In this example, the crude X1-Y association is confounded by U', but not by co-exposure X2. The crude X2-Y association is confounded by X1, but not by U' as that pathway is blocked by the collider X1.

The natural tendency is again to mutually adjust for X1 and X2, particularly if we don't know that U' exists. What happens? It turns out that the X1-Y association (adjusted for X2) can be <u>more</u> biased than the crude estimate (Weisskopf et al. 2018). Meanwhile, adjusting for X1 eliminates the confounding by co-exposure for the X2-Y association (by blocking that pathway), but leads to confounding by U' (opening the pathway through U' by conditioning on the collider X1).

We can quantify the results when the associations are linear and we apply linear regression, again assuming standardized variables. As previously shown (Weisskopf et al. 2018), the bias in the mutually adjusted X1-Y association is always amplified by $1/(1-\rho^2)$ compared to the crude result. The X2-Y association is also typically biased. For example, if all four coefficients ($c_1$, $c_2$, $c_3$, and $\rho=c_4c_5$ here) are positive, then the X2-Y association switches sign from positive for the crude association to negative for the mutually adjusted association. See Table 2.

## 2. Equivalency of data and regression results between different DAGs

The two examples illustrate the importance of having a causal model in mind when analyzing data and interpreting results. With the same three variables (X1, X2, Y), mutual adjustment eliminates bias in the confounding by co-exposure case, while mutual adjustment can make bias worse in the CAB case. Importantly, one cannot tell from the regression results which of these DAGs (or many others) is correct. Subject area specific information is required.

Indeed, the same underlying data can be consistent with both DAGs, i.e., both DAGs can generate exactly the same crude and adjusted regression results. Comparison of Tables 1 and 2 show the correspondence between parameters. Equating the adjusted regression results,

$$\beta_{1|2} = b_1 = c_1 + \frac{c_2 c_3}{1-\rho^2} \tag{1a}$$

$$\beta_{2|1} = b_2 = -c_2 c_3 \left(\frac{\rho}{1-\rho^2}\right) \tag{1b}$$

we directly obtain the causal coefficients for figure 1a ($b_1$, $b_2$) from those for figure 1b ($c_1$, $c_2$, $c_3$). Some algebra (not shown) allows one to express ($c_1$, $c_2c_3$) as a function of ($b_1$, $b_2$); note that only the product $c_2c_3$ is determined (and not $c_2$ and $c_3$ individually).

$$c_1 = b_1 + \frac{b_2}{\rho} \tag{2a}$$

$$c_2 c_3 = -b_2 \left( \frac{1 - \rho^2}{\rho} \right) \tag{2b}$$

This means that the regression result $\beta_1$ doesn't tell us whether we are obtaining the true causal estimate that we want ($b_1$) under DAG 1a, or a biased estimate ($c_1 + c_2c_3/(1 - \rho^2)$) of the causal estimate we want ($c_1$), under DAG 1b, and similarly for $\beta_2$.

The same correspondences (equations 1,2) are obtained using the crude results. (To derive equivalent parameters between DAGs, the unit variance assumptions may sometimes need to be relaxed).

There are of course other DAGs consistent with two exposure variables and one outcome variable, many more if one allows the number of unknown or unmeasured variables to increase. Clearly, regression cannot in and of itself tell us the correct underlying DAG. We cannot determine from data alone the best way to analyze data or interpret results. This is true for both real world data and synthetic data.

## 3. Generation of synthetic mixtures epidemiology data

Collinearity and confounding by co-exposure are two of the issues frequently mentioned in mixtures epidemiology. For example, in Figure 1a, we might wonder about the effect of high correlation between X1 and X2 on bias in effect estimates or selection of variables. For example, if $b_2=0$, how would our ability to select only X1 be affected by increasing $\rho$? In some situations and for some statistical methods we can answer these questions mathematically. In other cases, we would typically generate synthetic data and compare the results with the true model used to generate the data. But in this case, we must pay attention to the way we simulate the data.

Suppose we want to generate synthetic data corresponding to Fig 1a, confounding by co-exposure, for a linear model with parameters $b_1$, $b_2$, and $\rho=b_3 b_4$. Let's further assume, for simplicity, that X1 and X2 are standard normal (these assumptions can be relaxed if desired, e.g., assuming log-normal data or correlated quantile data, e.g., Keil et al 2020). At least two methods have been used.

3a. Method 1:

Follow these steps to simulate data for a linear model for the scenario described by Figure 1a, building directly from the DAG:

1) Pick values for the causal coefficients $b_1$, $b_2$, $b_3$, $b_4$ as well as the sample size n.

2) Generate $n$ values of U assuming a standard normal distribution.

3) Generate X1 and X2 using the following equations:

$$X1 = b_3 U + \delta_1$$
$$X2 = b_4 U + \delta_2$$
(3)

where $\delta_1$ and $\delta_2$ are error terms. In terms of DAGs, this is equivalent to adding new variables $\delta_1$ and $\delta_2$ that point only into X1 and X2, respectively, as in Fig 1c; such variables are typically omitted in DAGs because they are not a cause of more than one variable (and therefore cannot introduce structural bias). As discussed in the Supplemental Material, we include these error terms (with appropriate variances) so that X1 and X2 are standard normal with $\rho=b_3 b_4$ equal to their correlation coefficient.

4) Generate the expected values of Y using the following equation:

$$E[Y] = b_1 X_1 + b_2 X_2$$
(4)

The two terms correspond to the two arrows pointing directly into Y from X1 and X2. For simplicity, we omit the constant term in (4).

5) Obtain the vector Y of outcome data with some added noise, by adding to (4) a vector of normal error terms $\varepsilon$ (centered at zero) with variance $\sigma^2$:

$$Y = E[Y] + \varepsilon = b_1 X_1 + b_2 X_2 + \varepsilon$$
(5)

Again, this corresponds to a new variable $\varepsilon$ pointing into Y (see Fig 1c). If we also want Y to have a specified variance, e.g., $V(Y)=1$, then we apply the standard equation for the variance of sums to (5) and rearrange, obtaining

$$V(\varepsilon) = V(Y) - b_1^2 - b_2^2 - 2\rho b_1 b_2$$
(6)

Since variances must be non-negative, this places constraints on the parameters $\rho$, $b_1$, $b_2$. Note that $V(\varepsilon)$ depends on $\rho$.

<u>3b. Method 2:</u>

There is a second approach that can sometimes simplify the simulation (e.g., Carrico et al 2015, Czarnota et al 2015). Suppose we want to simulate data for the DAG in figure 1a, assuming that X1 and X2 are multivariate normal with correlation $\rho$ and the underlying linear model for Y as in (4). Then Y is also normal, i.e., (X1, X2, Y) are multivariate normal. With this insight, one can generate a set of synthetic data for Figure 1a as follows:

1) Pick values for $\rho$, $b_1$, $b_2$ as well as the sample size n.

2) Compute the crude correlation coefficients between the three variables. The correlation coefficient for X1-X2 is of course $\rho$. When variances equal one, the correlations for X1-Y and X2-Y are equal to the crude coefficients in Table 1: $r_{1y} = b_1 + \rho b_2$ and $r_{2y} = b_2 + \rho b_1$.

3) Create the symmetric variance-covariance matrix for the three variables (X1, X2, Y):

$$\Sigma = \begin{bmatrix} 1 & \rho & r_{1y} \\ \rho & 1 & r_{2y} \\ r_{1y} & r_{2y} & 1 \end{bmatrix} \qquad (7)$$

We have assumed unit variances for X1, X2 and Y, but this can be relaxed. Note that $\Sigma$ must be positive definite.

4) Generate the multivariate normal data (X1, X2, Y) directly using (7), for example by using the mvrnorm function in R.

<u>3c. Hybrid methods</u>

In practice, a third approach for creating synthetic data is often used that combines elements of both methods 1 and 2. In our example for method 1 above, we first created U and then built X1 and X2 from that, including the extra error terms $\delta_1$ and $\delta_2$. This extra effort is not necessary if the goal is to obtain X1 and X2 as bivariate normal. Hence, one might use a method 2 approach to simulate X1 and X2 and then a method 1 approach to simulate Y. Our opinion is that it doesn't really matter as long as one is clear about the DAG being simulated and the additional assumptions needed for turning a qualitative DAG into a data set. To guide construction of simulations and interpretation of results we recommend researchers always include a DAG and explicitly indicating with reference to the DAG which parameter(s) they are attempting to recover (as the "truth") with their mixtures analysis measure.

3d. Calculating bias using synthetic data

To numerically test a statistical method, e.g., linear regression, one would typically generate a data set (using one of these method), run linear regression—perhaps both crude and mutually adjusted—save the results, and repeat, say 1000 times. To calculate bias of the individual regression estimates assuming figure 1a, average the X1-Y estimates of the 1000 data sets and subtract the true value of $b_1$, and similarly for X2-Y using $b_2$. The variance of the regression coefficients can also be computed.

Methods 1 and 2 can also be used to generate data for figure 1b. Here one would pick parameters $c_1$, $c_2 c_3$, and n. A method 1 approach might first simulate U and U', use these results to simulate X1 and X2, and finally construct Y. A method 2 approach would use the crude coefficients from Table 2 for $r_{1y}$ and $r_{2y}$. To calculate bias of the regression estimates for Figure 1b, average the X1-Y estimates of the simulated data sets and subtract $c_1$, and similarly for X2-Y, where the true value is zero.

Note that method 1 is more flexible and can accommodate nonlinear functions. Method 2 is often simpler when the required assumptions are met, often allowing one to skip construction of extra variables, e.g., U, U'. But for a single set of parameters and a linear model, both methods can in principle generate exactly the same data, data consistent with either Figure 1a or 1b.

## 4. How is a given mixtures epidemiology method affected by increasing collinearity? How should we create synthetic data to answer this question?

We can now address the question of generating synthetic data to examine collinearity. The general topic of the pattern of correlations found in exposure data is beyond the scope of this paper (Weisskopf et al. 2018, Webster 2018). For our purposes here, it is sufficient to note that correlations between exposures typically vary from zero to nearly one and tend to form blocks of correlated variables; negative correlations can also occur but do not appear to be as common. The high degree of correlation shown by some exposures raises the spectre of collinearity, a standard issue for regression.

Hence, the effect of collinearity is a potentially important statistical issue for mixtures methods. Indeed, it is one of the motivations for some novel methods such as weighted quantile sum regression (WQS) (Carrico et al 2015). It is well known that in linear regression, high degrees of correlation do not lead to biased regression coefficients—biased in the statistical sense of difference from truth on average—but increase the size of standard errors and confidence intervals (e.g., Schisterman et al 2017). The effects of collinearity on novel mixtures methods are less well known and often less amenable to theoretical analysis.

Suppose we want to examine collinearity for the DAG in Figure 1a, confounding by co-exposure. More specifically, we want to determine the effect of increasing the correlation between X1 and X2 ($\rho$) while keeping the causal coefficients linking X1 to Y ($b_1$) and X2 to Y ($b_2$) fixed. We hypothesize that many researchers have this type of model in mind when investigating effects of collinearity, even if not explicitly stated.

We earlier described methods for simulating a linear model for Figure 1a. To simulate increasing collinearity, we might, for example, use the hybrid method to generate a series of data sets, keeping $b_1$, $b_2$ the same, while increasing $\rho$ (As discussed earlier, some care may be needed with respect to $V(\varepsilon)$ if variances will be examined). To examine bias, we would then compare the average X1-Y and X2-Y associations with $b_1$ and $b_2$, respectively.

As we saw earlier, methods 1, 2 and the hybrid approach can generate the same results for any single data set. One might therefore be tempted to use method 2, increasing $\rho$, but keeping $r_{1y}$ and $r_{2y}$ the same. However, recall that for the DAG in Figure 1a (and corresponding Table 1), $r_{1y}$ is the <u>crude</u> association between X1 and Y, similarly for X2:

$$r_{1y} = b_1 + \rho b_2 \qquad\qquad\qquad\qquad\qquad\qquad (7a)$$

$$r_{2y} = b_2 + \rho b_1 \qquad\qquad\qquad\qquad\qquad\qquad (7b)$$

These equations tell us that for fixed $b_1$ and $b_2$, changing $\rho$ will change $r_{1y}$ and $r_{2y}$. Put another way, changing $\rho$ while fixing $r_{1y}$ and $r_{2y}$ means that $b_1$ and $b_2$ have changed from those initially specified in step 1 of Method 2 above. But then comparing regression results with the original $b_1$ and $b_2$ to examine bias doesn't make sense. The problem is that we have changed all of the causal coefficients in Figure 1a ($b_1$, $b_2$ and $\rho=b_3b_4$) rather than just those governing the X1-X2 correlation (or we've abandoned the DAG in Figure 1a—see below). In fact, in order to keep the $r_{1y}$ and $r_{2y}$ fixed while changing $\rho$, one of $b_1$ or $b_2$ will move higher and the other lower—a phenomenon referred to as the reversal paradox that is often misinterpreted as a statistical effect on regression beta estimates when putting correlated variables in a model together rather than a consequence of having (usually unwittingly) changed the underlying causal parameters being modeled (e.g., Tu et al 2008, Vatcheva et al 2016). A similar problem would be associated with the DAG in Fig 1b. Suppose we want to fix $c_1$, $c_2$, $c_3$, and only change $\rho$ (=$c_4c_5$). Then, as shown in Table 2, $r_{1y}$ and $r_{2y}$ (which are the equations in the crude column) change.

  Figures 2a and 2b show two DAGs where changing the causal coefficients governing the X1-X2 correlation <u>does</u> allow $r_{1y}$ and $r_{2y}$ to remain fixed (there are other possibilities, e.g., a hybrid of figures 1a and 2a). Examination of figure 2a, nicknamed "three handles" because there are three unknown variables (U, U', U''), shows that $r_{1y}$ (=$c_3c_4$) is insulated from changes in $\rho$ (=$c_1c_2$) by the colliders X1 and X2, and similarly for $r_{2y}$ (=$c_5c_6$). Crude analysis of the X1-Y and X2-Y associations using linear regression yields $c_3c_4$ and $c_5c_6$, respectively, and can be seen to be independent of $\rho$ (=$c_1c_2$). A mixtures method that recovers these values would, in a certain sense, be recovering the correct values, although they do not reflect causal X1-Y and X2-Y associations because they are actually due to confounding by U' and U''. Mutual adjustment produces other non-causal estimates (Table 3) as it conditions on the colliders X1 and X2. For example, for a crude analysis of X1-Y, the only open pathway is X1←U'→Y. Adjusting for X2 opens up the additional pathway X1←U→X2←U''→Y. Similar logic applies for analysis of X2.

An alternative DAG that would allow for $\rho$ to be changed with the r remaining fixed is one of reverse causation (Figure 2b). In this case, the only open pathway in a crude analysis of the X1-Y association is X1←Y since X2 is a collider, and vice versa. Again, assuming linear models and unit variance, the DAG shows that

$$r_{1y} = c_3 \tag{8a}$$
$$r_{2y} = c_4 \tag{8b}$$
$$\rho = c_1c_2 + c_3c_4 \tag{8c}$$

where the $c_i$ are the causal coefficients in Figure 2b and $\rho$ is the X1-X2 correlation. Equation (8) shows that we can change $\rho$ (by changing $c_1c_2$) while keeping other causal coefficients and, therefore, $r_{1y}$ and $r_{2y}$ fixed. In this example, crude analysis of the X1-Y and X2-Y associations yields $c_3$ and $c_4$, but they reflect reverse causality. Mutual adjustment conditions on colliders X1 and X2 and so introduces a non-causal pathway into estimation of each parameter (not shown).

Summing up, using method 2 to generate synthetic data by increasing $\rho=b_3b_4$ and fixing the $r_{iy}$ means that we are either changing other causal coefficients ($b_1$, $b_2$) in Figure 1a, or that we are using a different DAG (e.g., figure 2a). A possible example of this is the evaluation of WQS and other methods by Carrico *et al.* (2015). While the simulation scenarios include more than two exposures, they involve changing the correlations between exposures while fixing the $r_{iy}$ (with some equal to zero) or changing the $r_{iy}$ while fixing the correlations between exposures—a scenario that is possible under a DAG like that in figure 2a, but not 1a. Rather than calculate bias, they determine sensitivity and specificity of detecting an association with the outcome, treating non-zero $r_{iy}$ as truth. However, for the DAG in figure 2a, crude regression analysis provides the "correct" answers – i.e., the crude $r_{iy}$, which, however, are non-causal associations – while adjusted regression analysis (i.e., including all exposures in the model as WQS does) estimates results different from the original $r_{iy}$ with additional bias due to conditioning on colliders. Indeed, Carrico *et al.* found that crude ordinary regression had good sensitivity and specificity when the $r_{iy}$ were large, which is exactly what one would expect. This simulation procedure is not consistent with the confounding by co-exposure DAG (Figure 1a, where adjusted analyses should provide the correct answer), unless the underlying causal coefficients linking exposure and outcomes ($b_i$) change. In contrast, method 1 or the hybrid method makes it straightforward to fix $b_1$ and $b_2$ in Figure 1a while increasing ρ.

    None of this means that method 2 for generating synthetic data is wrong. It does, however, mean that it is important to have a DAG explicitly in mind when generating synthetic data and interpreting regression results in terms of whether they are reflecting the causal parameters we hoped to estimate or confounded

associations. This is particularly important when examining the effects of increased collinearity between exposures.

## 5. Are more sophisticated mixtures methods susceptible to co-exposure amplification bias (CAB)?

As discussed in our earlier paper (Weisskopf et al. 2018), it is an open question whether mixtures methods other than linear regression are also subject to co-exposure amplification bias (CAB).  For many such methods, one cannot easily write closed form solutions akin to Tables 1-3. The standard approach to answering this question would be to create synthetic data based on the DAG for CAB and test the methods.

But there is another conceptual approach based on the fact that different DAGs can generate the same data. This correspondence of DAGs means that any mixtures method that can replicate the results of linear regression for individual components in figure 1a (mutual confounding by co-exposure) will also be subject to CAB (figure 1b).

For example, with small amounts of error and the right degree of smoothing, smoothing methods (e.g., splines, generalized additive models) can very closely reproduce linear regression of linear models. Even though smoothing methods don't directly provide regression coefficients, the smooths then closely approximate linear regression and the underlying model. Thus, bkmr and similar smoothing approaches should also be subject to CAB.

The recently proposed quantile g-computation (qgcomp) method (Keil et al 2020) is a more interesting case. We will restrict our discussion to the situations we've discussed in this paper: the DAGs we've described with underlying linear models and no effect measure modification or interactions and time-fixed exposures. The underlying causal method, g-computation, should then yield beta estimates identical to linear regression. Indeed, it would typically use a linear regression model as the first step. Thus, g-computation should be susceptible to CAB. qgcomp converts data into quantiles before analysis, but on already quantiled data, g computation should again yield the same beta coefficients for individual exposures as linear regression.

One difference in qgcomp is the estimation of the overall effect of the mixture (it also computes weights for components). For the confounding by co-exposure DAG (fig. 1a) with the linear model of equation (5), it should on expectation produce the overall causal effect of the mixture

$$\psi = b_1 + b_2 \tag{9}$$

i.e., the sum of the underlying causal coefficients (Keil et al 2020). Notably, qgcomp does not assume that $b_1$ and $b_2$ have the same sign. For the CAB DAG (fig. 1b), or the reparameterization of 1b in terms of 1a, the true value of $\psi$ is just $c_1$. As U' is unknown and therefore omitted (violating the assumption of the causal model), the expected overall effect is found by adding the beta coefficients in the right column of Table 2:

$$\psi = \left(c_1 + \frac{c_2 c_3}{1 - \rho^2}\right) + \left(-c_2 c_3 \frac{\rho}{1 - \rho^2}\right) = c_1 + \frac{c_2 c_3}{1 + \rho} \tag{10}$$

For no correlation between exposures ($\rho=0$), we obtain $c_1+c_2c_3$ (since then U disappears in Fig. 1b) but the confounding by U' ($c_1c_2$) remains. The bias of the total effect of the mixture $\psi$ is just equation 10 minus the true causal value $c_1$:

$$\psi bias = \frac{c_2 c_3}{1+\rho} \tag{11}$$

For positive correlations between exposures, the absolute value of the bias of $\psi$ decreases as we increase $\rho$. This occurs because even though each individual beta coefficient becomes more biased (due to CAB), the two beta coefficients have opposite signs and partly cancel each other out. Not only is there no CAB for the total effect of the mixture ($\psi$), but the bias from the uncontrolled confounding (by U' in Figure 1b) is reduced as correlation increases. The same conclusion holds true for the estimate of $\psi$ derived from linear regression by adding the two beta coefficients (Note that qgcomp has other useful properties compared to linear regression (Keil et al 2020)). However, for negative correlations between exposures, the absolute value of the bias of $\psi$, calculated using either method, increases as $\rho$ becomes more negative. Estimates of the total effect of a mixture may have advantages, especially since negative correlations between exposures appear to be rarer than positive correlations (Webster 2018). More research is needed on the susceptibility of other mixtures methods such as WQS to CAB as well as the properties of measures of total effect.

## 6. Conclusion

In summary, when results from a given analysis are considered, precise assumptions about the underlying data structure—which are often represented in a DAG—must be made when trying to make causal inferences from the results. While analysis results can rule out some DAGs, a number of DAGs can be consistent with the same results. Thus, one cannot infer the DAG from the data. This also has implications for the generation of synthetic data, which are very useful for testing the capabilities of new mixtures methods. There are a number of ways that such synthetic data may be created, but we strongly recommend that researchers explicitly use a DAG when doing so. Finally, we use the equivalency of DAGs to provide a test of whether novel mixtures methods are subject to co-exposure amplification bias. Our results indicate that bkmr is susceptible to this problem, while estimates of the total effect of a mixture, obtained via linear regression or gqcomp, may sometimes avoid it. Estimates of the total effect of a mixture is an important but relatively underexplored topic that warrants further investigation.

**Acknowledgements:**

This work was supported by NIEHH grant R01ES028800.

**Table 1.** Linear regression results for DAG 1a: confounding by co-exposure ($\rho = b_3 b_4$)

| Association | Crude ($r_{iy}$) | Mutually adjusted |
|---|---|---|
| $\beta_1$ (for X1-Y) | $b_1 + \rho b_2$ | $b_1$ |
| $\beta_2$ (for X2-Y) | $b_2 + \rho b_1$ | $b_2$ |

**Table 2.** Linear regression results for DAG 1b: co-exposure amplification bias ($\rho = b_3 b_4$)

| Association | Crude ($r_{iy}$) | Mutually adjusted |
|---|---|---|
| $\beta_1$ (for X1-Y) | $c_1 + c_2 c_3$ | $c_1 + c_2 c_3 / (1 - \rho^2)$ |
| $\beta_2$ (for X2-Y) | $\rho c_1$ | $-c_2 c_3 \rho / (1 - \rho^2)$ |

**Table 3.** Linear regression results for DAG 2a "3 handles" ($\rho = c_1 c_2$)

| Association | Crude ($r_{iy}$) | Mutually adjusted |
|---|---|---|
| $\beta_1$ (for X1-Y) | $c_3 c_4$ | $(c_3 c_4 - \rho c_5 c_6)/(1 - \rho^2)$ |
| $\beta_2$ (for X2-Y) | $c_5 c_6$ | $(c_5 c_6 - \rho c_3 c_4)/(1 - \rho^2)$ |

**Figure 1**. A) DAG for confounding by co-exposure with correlation between exposures $\rho=b_3b_4$. B) DAG for co-exposure amplification bias, $\rho=c_4c_5$. C) DAG for confounding by co-exposure with noise (errors) explicitly shown, $\rho=b_3b_4$.

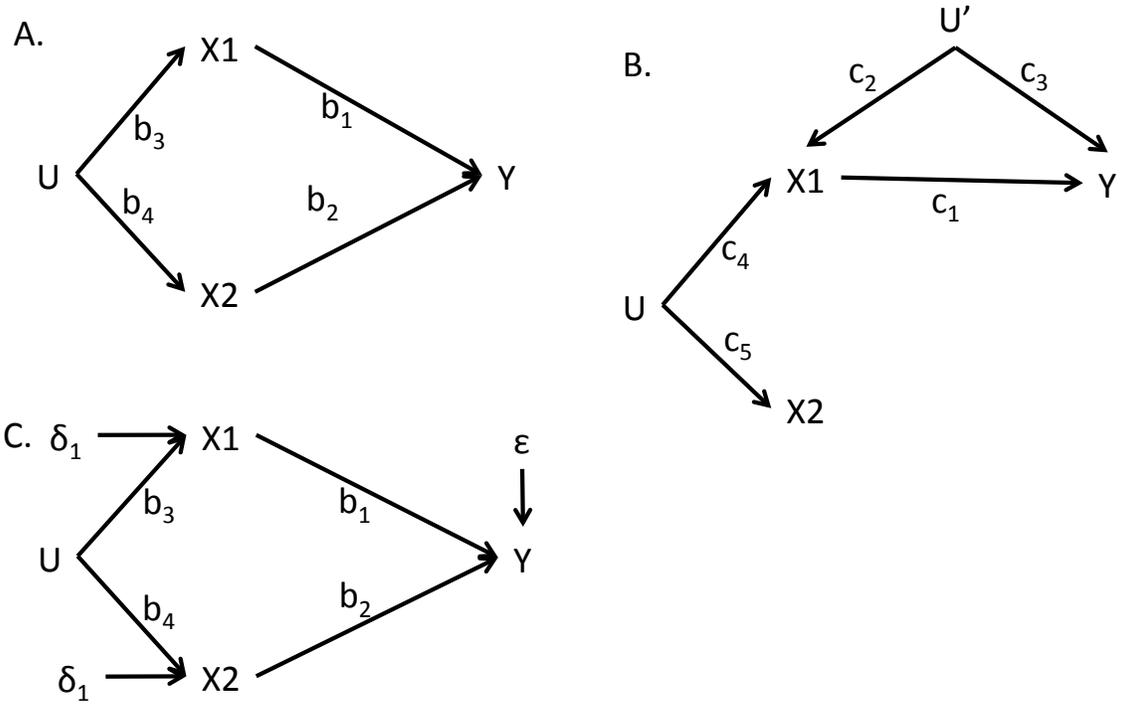

**Figure 2**. Two DAGs that allow the X1-X2 correlation to be change while fixing the X1-Y and X2-Y correlations: A) "3 handles" ($\rho=c_1c_2$, $r_{1y}=c_3c_4$, $r_{2y}=c_5c_6$ ). B) Reverse causation ($\rho=c_1c_2+c_3c_4$, $r_{1y}=c_3$, $r_{2y}=c_4$). There are other possibilities, e.g., adding to Figure 2a the causal X1-Y and X2-Y links from Figure 1a.

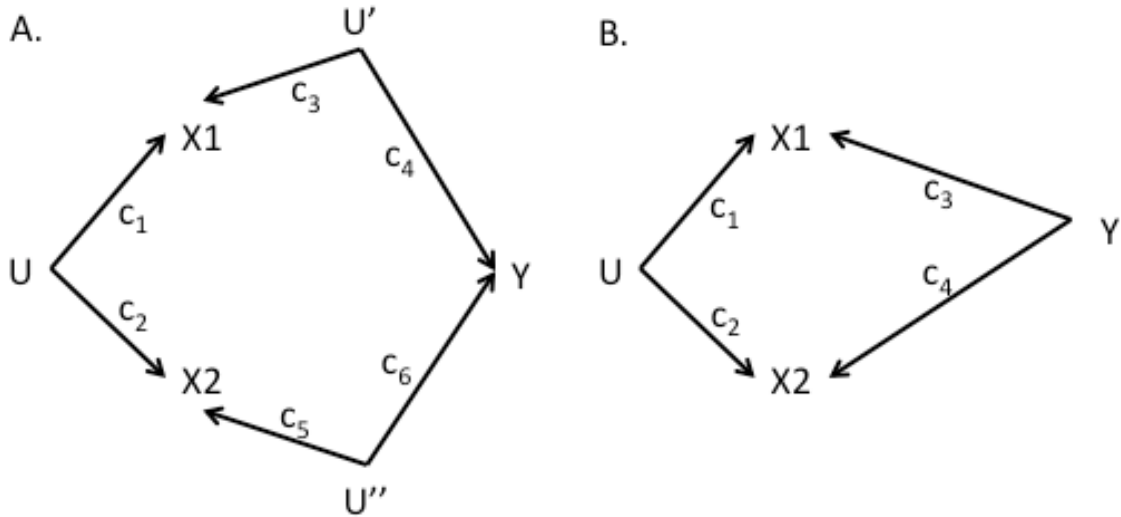